# Treating reflective indicators as causal-formative indicators in order to compute factor score estimates or unit-weighted scales

André Beauducel[1], Anja Leue[2] & Norbert Hilger[1]

1 University of Bonn, Germany, Institute of Psychology
2 University of Kiel, Germany, Institute of Psychology

**Abstract**

Individual scores on common factors are required in some applied settings (e.g., business and marketing settings). Common factors are based on reflective indicators, but their scores cannot unambiguously be determined. Therefore, factor score estimates and unit-weighted scales are used in order to provide individual scores. It is shown that these scores are based on treating the reflective indicators as if they were causal-formative indicators. This modification of the causal status of the indicators should be justified. Therefore, the fit of the models implied by factor score estimates and unit-weighted scales should be investigated in order to ascertain the validity of the scores.

**Introduction**

Most psychometric models assume that latent variables are based on reflective (effect) indicators. This assumption means that the latent variables (true scores) determine the indicators (observed scores or item scores). This perspective is taken in classical test theory (Lord & Novick, 1968), in factor analysis (Harman, 1976), in conventional structural equation modelling (Bollen, 1989), as well as in item response theory (Kohli, Koran, & Henn, 2015; Meade & Lautenschlager, 2004). The causal perspective is reversed when causal-formative indicators are considered: Causal-formative indicators are assumed to directly affect their corresponding latent variable (Blalock, 1964; Bollen & Bauldry, 2011). There have been several critical statements on formative indicators (e.g., Edwards, 2011; Iacobucci, 2010; Lee, Cadogan, & Chamberlain, 2013). Bollen and Diamantopoulos (2015) examined several of these criticisms and concluded that they are partly related to a confusion between different versions of formative indicators. It follows from their analysis that at least causal-formative indicators may be of interest for further research.

Moreover, in business settings, the "usability of a software product" is defined in the international norm ISO 9241 by three criteria such as effectiveness, efficiency, and satisfaction (Christophersen & Konradt, 2008). These criteria are of special importance to evaluate the quality of user interfaces and E-commerce sites. In the same line, "principles of dialogue design" incorporate further criteria that might be conceived as formative indicators (e.g., suitability for the task, self-descriptiveness, controllability conformity with user expectations, error tolerance, suitability for individualization, suitability for learning). Moreover, Coltman, Devinney, Midgley, and Venaik (2007) argued that "integration responsiveness" and "market orientation" also constitute formative indicators in international business and marketing disciplines. Petter, Straub and Rai (2007) suggest "organizational performance" to be a construct built by means of formative indicators such as productivity, profitability and market share. Each indicator operationalizes a specific aspect of the construct "organizational performance".

Usually, the constructs (e.g., usability of software, quality of user interfaces, organizational performance) that are defined by formative indicators have in common that less is known about their inter-correlations. Consequently, factor analysis and estimation of Cronbach's alpha reveal less appropriate methods for the investigation of validity and reliability of formative indicators (Christophersen & Konradt, 2008). Moreover, it is noteworthy that changes of the formative indicators that comprise a construct inherently cause changes in the meaning and the content of the construct (Petter et al., 2007). To sum up, there are several applied contexts, in which formative indicators are regularly used. In addition to the examples presented here, it will be shown below that any applied setting that requires individual scores will be based on treating at least some indicators of latent variables as formative indicators.

In Table 1 we show the causal-formative indicators of latent variables as they have been defined by Bollen and Diamantopoulos (2015, p. 4, Table 2) in Equations 1-3. The $p$ indicators ($x_i$) are assumed to measure the same concept and act as measures of a latent variable ($\eta$) representing the concept. Assuming that the indicators do not completely determine the latent variable, a disturbance term ($\zeta$) is introduced in Equation 1, whereas a zero disturbance is assumed in Equations 2 and 3. In Equations 1 and 2 the weights ($\beta_i$) are estimated, whereas in Equation 3, the weights $\beta_i^*$ are fixed. Equations 1-3 do not tell whether it is justified to treat the indicators ($x_i$) as causal-formative indicators. This justification is important and can only be inferred from theory (Bollen & Diamantopoulos, 2015). Irrespective of the justification, the indicators are in fact treated as causal-formative indicators in Equations 1-3 because the linear combinations of the indicators (plus an error term in Equation 1) yield the latent variables. Bollen and Davis (2009) have shown how to specify models with causal-formative indicators within structural equation modeling (SEM). In the following, we consider models that are initially based on reflective indicators and what happens to this perspective when individual scores are calculated.

Table 1.
Causal-formative indicator variables

| | | | |
|---|---|---|---|
| Indicator | weights | $\eta = \beta_1 x_1 + \beta_2 x_2 + ... + \beta_p x_p + \zeta$ | (1) |
| estimated | | $\eta = \beta_1 x_1 + \beta_2 x_2 + ... + \beta_p x_p$ | (2) |
| Indicator | weights | $\eta = \beta_1^* x_1 + \beta_2^* x_2 + ... + \beta_p^* x_p$ | (3) |
| prespecified | | | |

*Notes.* "*" indicates that the weights were fixed (not estimated). Adapted from Bollen and Diamantopoulos (2015, p. 4, Table 2)

**Factor score estimates and causal-formative indicators**

The scores of latent variables that are based on causal-formative indicators are, of course, directly determined by the indicators. In contrast, the scores of common factors that are based on reflective indicators cannot unambiguously be defined (McDonald & Burr, 1967). The reason is that the number of common and unique factors exceeds the number of (reflective) indicators. This follows from the definition of the common factor model. In the population, the common factor model is defined as

$$\mathbf{x} = \mathbf{\Lambda}\mathbf{f} + \mathbf{e}, \tag{4}$$

where $\mathbf{x}$ is the random vector of indicators of order $p$ and $\mathbf{f}$ is the random vector of common factors of order $q$. Moreover, $\mathbf{e}$ is the random vector of error factors of order $p$, and $\mathbf{\Lambda}$ is the matrix of common factor loadings of order $p \times q$. The common factors $\mathbf{f}$ and the error factors $\mathbf{e}$ are assumed to have an expectation zero ($\varepsilon[\mathbf{x}] = 0$, $\varepsilon[\mathbf{f}] = 0$, $\varepsilon[\mathbf{e}] = 0$). The covariance between the common factors and the error factors is assumed to be zero ($\mathrm{Cov}[\mathbf{f}, \mathbf{e}] = 0$). That the factor model is based on reflective indicators follows from Equation 4. However, Equation 4 does not imply that there is a theoretical justification for considering the indicators as reflective indicators. The justification of treating the observed variables as reflective indicators has to be provided from theory. The covariance matrix of reflective indicators $\mathbf{\Sigma}$ can be decomposed into

$$\mathbf{\Sigma} = \mathbf{\Lambda \Phi \Lambda'} + \mathbf{\Psi^2}, \tag{5}$$

where $\mathbf{\Phi}$ represents the $q \times q$ factor correlation matrix and $\mathbf{\Psi^2}$ is a $p \times p$ diagonal matrix representing the expected covariance of the error factors $\mathbf{e}$ ($\mathrm{Cov}[\mathbf{e},\mathbf{e}] = \mathbf{\Psi^2}$). It is assumed that $\mathbf{\Psi^2}$ is positive definite and that the expectation of the non-diagonal elements is zero.

Since there are $q$ common factors and $p$ error factors, there are $p + q$ factors, whereas there are only $p$ variables. It is therefore impossible to determine the scores of the factors. This problem is referred to as factor score indeterminacy (Guttman, 1955). Nevertheless, scores on common factors may sometimes be of interest. For example, in the context of personnel selection it might be helpful to base the selection of individuals on the scores which individuals have on a latent variable of interest. It has therefore been proposed to compute factor score estimates as linear combinations of the indicators (e.g., Thurstone, 1935; Bartlett, 1937; McDonald, 1981). For $q = 1$, Thurstone's (1935) regression factor score estimator $\mathbf{f}_r$ can be written as

$$\mathbf{f}_r = \mathbf{B'x} = \beta_1 x_1 + \beta_2 x_2 + ... + \beta_p x_p, \tag{6}$$

with the weight-matrix $\mathbf{B} = \mathbf{\Sigma^{-1} \Lambda \Phi}$ being computed from the covariance matrix of the indicators ($\mathbf{\Sigma}$), the factor loading matrix ($\mathbf{\Lambda}$), and the inter-correlations of the factors ($\mathbf{\Phi}$). For $q > 1$ the matrix notation introduced in Equation 6 would be appropriate. Although the relationship between the common factor and the factor score estimates will be different for Bartlett's (1937) factor score estimator and for McDonald's (1981) factor score estimator, Equation 6 also describes the computation of these factor score estimators. Obviously, the right hand side of Equation 6 equals the right hand side of Equation 2. Thus, when scores on common factors that were initially based on reflective indicators are calculated, these scores will be treated as being based on causal-formative indicators. If there are relevant advantages of common factors as latent variables being based on reflective indicators over latent variables that are based on causal-formative indicators, these advantages are no longer present when real scores on the latent variables are computed. In other words, the possible advantages of using reflective indicators as a basis for common factors are related to the indeterminacy of the factors. There might be theoretical reasons for considering the indicators as being reflective.

Nevertheless, determinate factor score estimates can only be computed when the indicators are treated as causal-formative indicators. Moreover, if the theoretical justification for treating the indicators as reflective indicators in factor analysis was substantial, the factor score estimates do no longer correspond to this theoretical perspective. The impossibility to compute scores for common factors that preserve the reflective property of the indicators might be regarded as a major flaw of factor score estimates. However, resigning to the computation of individual scores of common factors in order to preserve the virtues of reflective indicators might not help in psychological assessment or in other applied settings, where individual scores are needed in order to prepare decisions on the individual level.

**Testing the model implied by factor score estimates**

Especially, in the context of confirmatory factor analysis models based on reflective indicators are tested. When a factor model fits to the data, this does not imply that a model that treats these indicators as being causal-formative, also fits to the data. When factor score estimates are computed, it might therefore be necessary to test whether the corresponding model, which is based on treating the indicators as causal-formative indicators, fits to the data. Thus, the model implied by the factor score estimates should be tested, even when a factor model based on reflective indicators has been tested before. A corresponding test can be based on the covariance matrix of the indicators ($\Sigma$) and on the covariance matrix of the indicators as it can be reproduced from the factor score estimates. According to Beauducel (2007), the covariance matrix of the indicators as it can be reproduced from most of the factor score estimates (comprising Thurstone's, Bartlett's, and McDonald's factor score estimate), is

$$\Sigma_{fs} = \Lambda(\Lambda'\Sigma^{-1}\Lambda)^{-1}\Lambda'. \qquad (7)$$

It is possible to evaluate the discrepancy between $\Sigma$ and $\Sigma_{fs}$ in order to investigate whether the model implied by a given factor score estimate fits to the data. One way to evaluate this discrepancy is the standardized root mean square residual (SRMR). The corresponding

SRMR can be computed as

$$\text{SRMR} = (\mathbf{1}'((\Sigma - \Sigma_{fs})^{\circ 2} + diag(\Sigma - \Sigma_{fs})^{\circ 2})\mathbf{1} \, p^{-1}(p+1)^{-1})^{1/2}, \qquad (8)$$

where $\mathbf{1}$ is a $p \times 1$ unit-vector and "$^{\circ 2}$" denotes the elementwise squares. There are, of course, other possibilities to evaluate model fit. At this point, it should only be noted that such an evaluation is relevant and possible.

**Unit-weighted scales and causal-formative indicators**

Unit-weighted scales are often regarded as a simple method for the computation of factor score estimates (DiStefano, Zhu, Mindrila, 2009). The computation of factor score estimates by means of unit-weighted sums of indicators has sometimes been recommended, even as a valuable alternative to more sophisticated procedures (Tabachnick, & Fidell, 2013). It has been recommended to base the unit-weighted scales on the sum of the indicators with salient loadings on a factor, but also as the sum of indicators with salient factor score coefficients (Grice, 2001; Grice & Harris, 1998). Even when there are different strategies for the identification of the relevant indicators, the procedure to compose unit-weighted sums of the indicators representing a factor has been regularly used in scale construction (Briggs & Cheek, 1986). When unit-weighted scales are computed for indicators that have previously been shown to load on a common factor, it is clear that they are assumed to measure the same concept. Since the weights of unit-weighted scales are fixed, they correspond to latent variables based on causal-formative indicators with fixed weights (Equation 3, Table 1). Thus, even when they are based on an initial factor analysis, the indicators of unit-weighted scales are not treated as reflective indicators when unit-weighted scales are computed. As for the factor score estimates the only way to avoid the use of causal-formative indicators is to avoid the use of unit-weighted scales.

**Testing the model implied by unit-weighted scales**

As noted above, it might be insufficient to test models with reflective indicators as a basis for the computation of scores that are based on treating the indicators as being causal-formative indicators. It might therefore be reasonable to test directly the model implied by unit-weighted scales, in which the indicators are treated as causal-formative indicators. It has already been proposed to compare the covariance matrix of the indicators with the covariance matrix of the indicators that is reproduced from the model implied by unit-weighted scales (Beauducel & Leue, 2013). Although Beauducel and Leue (2013) showed how to enter the parameters of unit-weighted scales into SEM-software, they did not consider explicitly the case of causal-formative indicators in their models. Moreover, they did not provide an algebraic framework for this approach. Therefore, the relevant algebraic framework is presented in the following.

Unit-weighted scales are

$$\mathbf{S_u} = \mathbf{B_u'}\mathbf{x},$$ (9)

where $\mathbf{x}$ is a random vector of $p$ indicators, whereas $\mathbf{B_u}$ contains the unit weights. When a single unit-weighted scale is investigated ($q = 1$), the weights $\mathbf{B_u}$ are a $p \times 1$ unit vector **1**. $\mathbf{B_u}$ contains a pattern of zeroes and ones when $q > 1$. According to Schönemann and Steiger (1976) the regression component loading pattern corresponding to the unit-weighted scales is

$$\mathbf{A_u} = \mathbf{\Sigma}\mathbf{B_u}(\mathbf{B_u'}\mathbf{\Sigma}\mathbf{B_u})^{-1},$$ (10)

where $\mathbf{\Sigma} = \varepsilon[\mathbf{xx'}]$ is the covariance matrix of indicators and $\mathbf{C_u} = \mathbf{B_u'}\mathbf{\Sigma}\mathbf{B_u}$ is the covariance of the unit-weighted scales. The loadings in $\mathbf{A_u}$ are not standardized when the unit-weighted scales are not standardized. The standardized loadings in $\mathbf{A_u}$ can be computed as

$$\mathbf{A_u} = \mathbf{\Sigma B_u D^{-1/2}(D^{-1/2}B_u^{'}\Sigma B_u D^{-1/2})^{-1}}, \tag{11}$$

with $\mathbf{D} = diag(\mathbf{B_u^{'}\Sigma B_u})$ representing the variance of the unit-weighted scales. The covariance matrix reproduced from the regression component loadings and from the covariances of the unit-weighted scales is

$$\mathbf{\Sigma_u} = \mathbf{A_u C_u A_u^{'}} = \mathbf{\Sigma B_u (B_u^{'}\Sigma B_u)^{-1} B_u^{'}\Sigma}. \tag{12}$$

The discrepancy between $\mathbf{\Sigma}$ and $\mathbf{\Sigma_u}$ represents the fit of the regression component model corresponding to the unit-weighted scales. For example, $\mathbf{\Sigma_u}$ can be entered instead of $\mathbf{\Sigma}_{fs}$ into Equation 8 for the SRMR in order to evaluate the fit of the model implied by unit-weighted scales. The standardized loadings can be entered as model parameters into SEM software when $\mathbf{\Sigma}$ is a correlation matrix (Beauducel & Leue, 2013), so that the fit of the model implied by the unit-weighted scales can be calculated. For $q = 1$ we get $\mathbf{D} = \mathbf{1^{'}\Sigma 1}$ so that Equation (11) becomes $\mathbf{A_u} = \mathbf{\Sigma 1 D^{-1/2}} = \mathbf{xx^{'}1D^{-1/2}}$. Since $\mathbf{D^{-1/2}1^{'}x}$ is a z-standardized unit-weighted scale, it follows that $\mathbf{A_u}$ contains the (not part-whole corrected) item-total correlations for $diag(\mathbf{\Sigma}) = \mathbf{I}$. These item-total correlations can be entered as loading parameters of the unit-weighted scales model into SEM software. If $\mathbf{\Sigma}$ is a non-standardized covariance matrix, the non-standardized loadings (Equation 10) should be entered into the software.

### Unit-weighted scales and composite-formative indicators

Although the problem of treating reflective indicators as if they were causal-formative indicators is similar for factor score estimates and unit-weighted scales, there is an important difference between factor scores estimates and unit-weighted scales. It makes no sense to compute factor score estimates when there is no conceptual unity of the indicators, because most factor loadings will be close to zero when there is no conceptual unity. Thus, the factors will not represent a relevant amount of variance when there is no conceptual unity. In contrast, it is possible to compute unit-weighted scales without an underlying factor model and without conceptual unity of the indicators. Thus, unit-weighted scales may not necessarily represent a common factor, they may result from predefined weights in the case that all formative indicators are equally important. Therefore, unit-weighted scales can also be based on composite-formative indicators, which are not based on conceptual unity. Nevertheless, as Bollen and Diamantopoulos (2015) noted, the fact that parameters are fixed a priori does not mean that one should not investigate whether the model is consistent with the data. It would, of course, be reasonable to perform the abovementioned evaluation of model fit for the unit-weighted scales model even when the weights are fixed a priori and when no conceptual unity of the composite-formative indicators can be assumed.

Bollen and Diamantopoulos (2015) referred to the difference between causal-formative and composite-formative indicators. Bollen and Bauldry (2011) introduced a similar difference in order to eliminate misunderstandings in the criticism of formative indicators. Besides theoretical considerations that can be a basis for the evaluation of the conceptual unity of the indicators, it might also be possible to consider the empirical inter-correlation between the indicators as a measure of their conceptual unity. Considering the empirical inter-correlation of the indicators as a basis for their conceptual unity implies that conceptual unity can be a matter of degree. Substantial correlations may indicate high conceptual unity, moderate correlations indicate moderate conceptual unity, and low or zero correlations indicate missing conceptual unity. Considering that conceptual unity is a matter of degree, it appears interesting to investigate how the size of the inter-correlations between indicators and the number of indicators affects the fit of the unit-weighted scale model.

**Indicator inter-correlations and the fit of the unit-weighted scale model**

The algebraic investigation of the effect of the indicator inter-correlations on the fit of the unit-weighted scales model was based on the model of parallel measurements. This simplification allows for the additional investigation of the effect of the scale length ($p$) on the fit of the unit-weighted scale model. When all indicators are z-standardized parallel measurements, all non-diagonal elements of the inter-correlation matrix correspond to the same indicator inter-correlation $r$. For this simple condition, it is possible to describe algebraically the effect of $r$ and $p$ on model fit. As an example, this is performed here for the SRMR. For a single unit-weighted scale ($q = 1$) the indicator correlation matrix is

$$\mathbf{\Sigma} = \mathbf{1}r\mathbf{1}' + (1-r)\mathbf{I}, \tag{13}$$

where $\mathbf{1}$ represents a $p \times 1$ unit-vector and $\mathbf{I}$ is a $p \times p$ identity matrix. Inserting $\mathbf{B_u} = \mathbf{1}$ for ($q = 1$) and $\mathbf{1}r\mathbf{1}' + (1-r)\mathbf{I}$ for $\mathbf{\Sigma}$ into Equation (12) yields

$$\mathbf{\Sigma_u} = (\mathbf{1}r\mathbf{1}' + (1-r)\mathbf{I})\mathbf{1}(\mathbf{1}'(\mathbf{1}r\mathbf{1}' + (1-r)\mathbf{I})\mathbf{1})^{-1}\mathbf{1}'(\mathbf{1}r\mathbf{1}' + (1-r)\mathbf{I}), \tag{14}$$

which can be transformed into

$$\begin{aligned}\mathbf{\Sigma_u} &= (rp\mathbf{1} + (1-r)\mathbf{1})(rp^2 + p(1-r))^{-1}(rp\mathbf{1}' + (1-r)\mathbf{1}') \\ &= \mathbf{1}\mathbf{1}'(r + (1-r)p^{-1}).\end{aligned} \tag{15}$$

The squared numerator of the SRMR can be computed as

$$\mathrm{SRMR}^2 p(p+1) = \mathbf{1}'((\mathbf{\Sigma} - \mathbf{\Sigma_u})^{\circ 2} + diag(\mathbf{\Sigma} - \mathbf{\Sigma_u})^{\circ 2})\mathbf{1}. \tag{16}$$

It follows from Equation 15 that the resulting matrix of $(\mathbf{\Sigma} - \mathbf{\Sigma_u})^{\circ 2} + diag(\mathbf{\Sigma} - \mathbf{\Sigma_u})^{\circ 2}$ contains $p(p-1)$ non-diagonal elements given by $(r - (r + (1-r)p^{-1}))^2$ and $2p$ diagonal elements given by $(1 - (r + (1-r)p^{-1}))^2$ so that Equation (16) can be written

as

$$\mathbf{1}'((\boldsymbol{\Sigma} - \boldsymbol{\Sigma_u})^{\circ 2} + diag(\boldsymbol{\Sigma} - \boldsymbol{\Sigma_u})^{\circ 2})\mathbf{1}$$
$$= p(p-1)(r - (r + (1-r)p^{-1}))^2 + 2p(1 - (r + (1-r)p^{-1}))^2. \quad (17)$$

Accordingly, the SRMR can be written as

$$\text{SRMR} = ((p-1)(p+1)^{-1}(r - (r + (1-r)p^{-1}))^2 + 2(p+1)^{-1}(1 - (r + (1-r)p^{-1}))^2)^{1/2}. \quad (18)$$

It follows from Equation 18 and from $\lim_{p \to \infty}(p-1)(p+1)^{-1} = 1$, $\lim_{p \to \infty}(1-r)p^{-1} = 0$, and $\lim_{p \to \infty}(p+1)^{-1} = 0$ that $\lim_{p \to \infty} \text{SRMR} = 0$. Moreover, $r = 1$ implies SRMR = 0. For different values of $p$, the $r$ necessary for obtaining an SRMR of .06, .09, and .12 is presented in Figure 1.

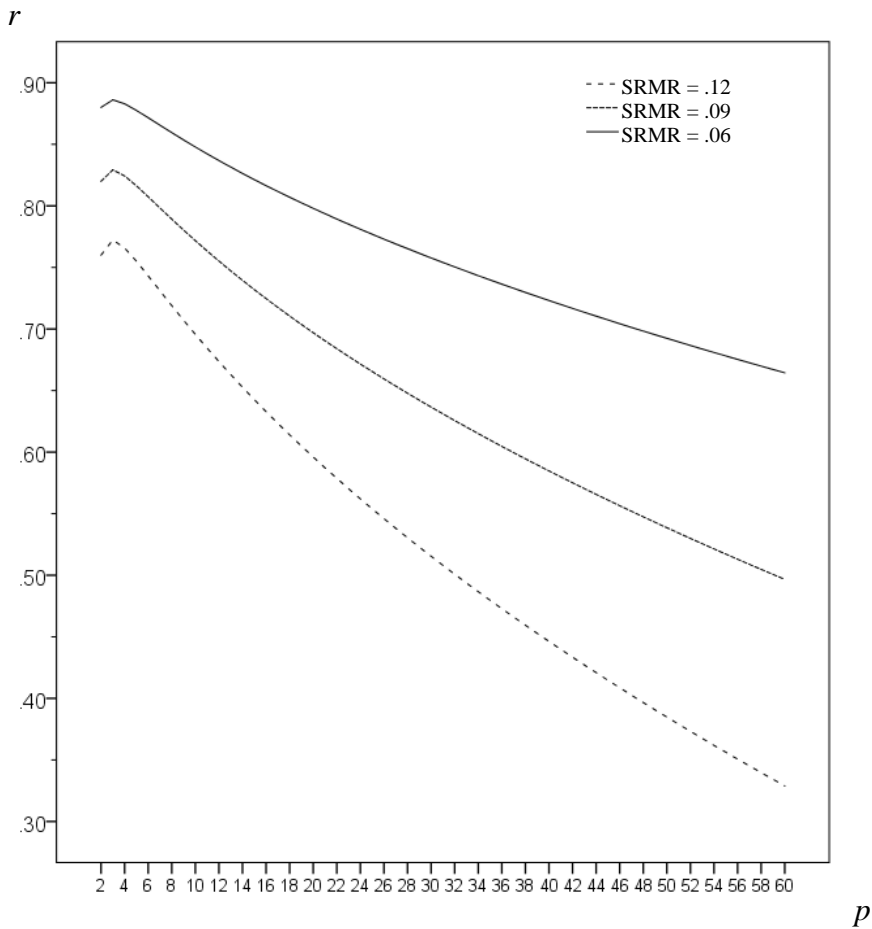

Figure 1. Relationship between $r$, $p$ and SRMR

These values for the SRMR cover the upper and the lower cutoff values that have been investigated by Hu and Bentler (1999). Using these cutoff values here does not imply that the current study advocates for the general use of cutoff values in the context of SEM fit indexes. It is rather likely that different cutoff values may be appropriate in different contexts (Marsh, Hau, & Wen, 2004). Figure 1 reveals that $r \geq .80$ is necessary for a 16 indicator unit-weighted scale in order to obtain an SRMR of .06 and that an $r$ of about .50 is necessary for a 60 indicator unit-weighted scale in order to obtain an SRMR of .09. This result indicates that the number of indicators has considerable consequences for the fit of the unit-weighted scales model. However, conceptual unity in terms of a high inter-correlation of the indicators is necessary to obtain a fit of the unit-weighted scales model when the scales have a length that is typical for psychological research and applications. It follows from Equation 18 that only for $p > 150$ indicators, a moderate fit of SRMR $= .09$ can be obtained with $r < .20$. Thus, conceptual unity of the indicators is a prerequisite of acceptable fit for unit-weighted scale models that are based on a realistic number of indicators. Of course, the results presented so far are limited to the population model of parallel measurements. A simulation study was therefore performed in order to investigate whether the abovementioned results generalize to unit-weighted scale models based on composite-formative indicators and whether the results generalize to the samples.

**Simulation Study**

The simulation study for the samples was based on one-factor population models with constant salient loadings (parallel measurements) and variable salient loadings (congeneric measurements). The mean salient loadings $l$ were .20, .40, .60, and .80 (corresponding to indicator inter-correlations of .04, .16, .36, and .64). The $p$ was 6, 12, and 24. Variable population loadings were generated by adding .10 to the mean salient loading for half of the indicators and by subtracting .10 from the mean salient loading for the other half of the indicators. From each population 5,000 samples were drawn and, for each sample, the SRMR$_S$ was calculated (the subscript "$s$" denotes computation for the samples).

The results of the simulation are summarized in Table 2, which contains the population values for SRMR as well as the means for the corresponding sample estimates (SRMR$_S$) for the constant as well as for the variable salient loadings. Overall, the results for the samples were extremely similar to the population results. Moreover, the results for the variable loadings sizes were nearly identical to the results for the constant loading sizes (see Table 1). Accordingly, the population results that have been obtained in the previous section can be generalized to the congeneric model.

Table 2.

Results for the SRMR for the simulations based on constant and variable loadings

| | | | | constant loadings | | | variable loadings | | |
|---|---|---|---|---|---|---|---|---|---|
| | | | | | SRMR$_S$ | | | SRMR$_S$ | |
| *n* | *L* | *r* | *P* | SRMR | *M* | *SD* | SRMR | *M* | *SD* |
| 150 | .2 | .04 | 6 | .45 | .45 | (.011) | .45 | .45 | (.011) |
| | | | 12 | .35 | .36 | (.005) | .35 | .36 | (.005) |
| | | | 24 | .26 | .27 | (.003) | .26 | .27 | (.003) |
| | .4 | .16 | 6 | .39 | .40 | (.015) | .39 | .40 | (.015) |
| | | | 12 | .31 | .31 | (.009) | .31 | .31 | (.008) |
| | | | 24 | .23 | .24 | (.005) | .23 | .24 | (.005) |
| | .6 | .36 | 6 | .30 | .30 | (.018) | .30 | .30 | (.017) |
| | | | 12 | .24 | .24 | (.012) | .24 | .24 | (.011) |
| | | | 24 | .18 | .18 | (.008) | .18 | .18 | (.008) |
| | .8 | .64 | 6 | .17 | .17 | (.015) | .18 | .18 | (.015) |
| | | | 12 | .13 | .14 | (.011) | .14 | .14 | (.010) |
| | | | 24 | .10 | .10 | (.008) | .11 | .11 | (.007) |
| 300 | .2 | .04 | 6 | .45 | .45 | (.008) | .45 | .45 | (.008) |
| | | | 12 | .35 | .36 | (.004) | .35 | .36 | (.004) |
| | | | 24 | .26 | .27 | (.002) | .26 | .27 | (.002) |
| | .4 | .16 | 6 | .39 | .39 | (.011) | .39 | .39 | (.011) |
| | | | 12 | .31 | .31 | (.006) | .31 | .31 | (.006) |
| | | | 24 | .23 | .23 | (.004) | .23 | .23 | (.004) |
| | .6 | .36 | 6 | .30 | .30 | (.012) | .30 | .30 | (.012) |
| | | | 12 | .24 | .24 | (.008) | .24 | .24 | (.008) |
| | | | 24 | .18 | .18 | (.006) | .18 | .18 | (.005) |
| | .8 | .64 | 6 | .17 | .17 | (.011) | .18 | .18 | (.011) |
| | | | 12 | .13 | .13 | (.008) | .14 | .14 | (.007) |
| | | | 24 | .10 | .10 | (.005) | .11 | .11 | (.005) |
| 900 | .2 | .04 | 6 | .45 | .45 | (.005) | .45 | .45 | (.005) |
| | | | 12 | .35 | .35 | (.002) | .35 | .35 | (.002) |
| | | | 24 | .26 | .26 | (.001) | .26 | .26 | (.001) |
| | .4 | .16 | 6 | .39 | .39 | (.006) | .39 | .39 | (.006) |
| | | | 12 | .31 | .31 | (.004) | .31 | .31 | (.003) |
| | | | 24 | .23 | .23 | (.002) | .23 | .23 | (.002) |
| | .6 | .36 | 6 | .30 | .30 | (.007) | .30 | .30 | (.007) |
| | | | 12 | .24 | .24 | (.005) | .24 | .24 | (.005) |
| | | | 24 | .18 | .18 | (.003) | .18 | .18 | (.003) |
| | .8 | .64 | 6 | .17 | .17 | (.006) | .18 | .18 | (.006) |
| | | | 12 | .13 | .13 | (.004) | .14 | .14 | (.004) |
| | | | 24 | .10 | .10 | (.003) | .11 | .11 | (.003) |

*Notes.* SRMR$_S$ denotes the SRMR for the samples, *n* = number of cases, *l* = salient loading size, *r* = item inter-correlations, *p* = number of variables.

**Empirical Example**

A sample of 191 students (99 females; Age: $M = 22.61$, $SD = 3.30$) from a German University filled in the German state version of the State-Trait-Anxiety-Inventory (STAI; Spielberger, Gorsuch, Lushene, Vagg, & Jacobs, 1983). All participants indicated written informed consent before the beginning of the study.

The STAI comprises 20 indicators with a 4-point Likert-type answer format and it had a high internal consistency (Cronbach's Alpha = .92). The correlation matrix of the indicators is given in the Appendix. The one-factor model based on z-transformed reflective indicators was tested by means of confirmatory factor analysis with maximum likelihood estimation (Mplus 7.11; Muthén & Muthén, 1998-2013). The factor model did not fit the data well according to conventional cutoff values (Hu & Bentler, 1998) for the Comparative Fit Index (CFI) ($\chi^2(170)$=467.48, $p < .001$, CFI = .831, SRMR = .067) with standardized factor loadings between .292 and .771 (see Appendix). Although the one-factor model is a suboptimal representation of the data according to the CFI, the SRMR might be regarded as sufficiently small. Now we consider that a researcher computes a unit-weighted scale for the STAI. The SRMR of the model based on unit-weighted formative indicators was .197 (calculated by means Equation 10; see Appendix for Syntax of SPSS Version 23). The SRMR of the model corresponding to the regression factor score estimator was .198 (calculated by means of Equations 7 and 8; see Appendix for SPSS-Syntax). Thus, although the SRMR was acceptable for the one-factor model, the corresponding SRMR for unit-weighted scales and factor score estimates was not acceptable. Changing the causal status of the variables from reflective to causal-formative resulted in a substantial decrease of model fit.

**Discussion**

When a factor model, which is based on reflective indicators is calculated, the factor scores are indeterminate. This implies that it is impossible to define these scores because the number of common and unique factors in the factor model is larger than the number of indicators. Therefore, individual factor score estimates can only be calculated when the latent variable is a linear combination of the indicators, which implies that the indicators are treated as causal-formative indicators of the latent variable. The same argument was presented with respect to unit-weighted scales. This implies that any criticism against causal-formative indicators should also be applied to factor score estimates and unit-weighted scales. An additional problem occurs when indicators that were considered as reflective indicators for some theoretical reason are treated as if they were causal-formative indicators. Using these indicators for the computation of factor score estimates or unit-weighted scales results in an inconsistent causal status of the indicators.

Moreover, we emphasized that the fit of models based on causal-formative indicators should be evaluated when factor score estimates or unit-weighted scales are calculated as a basis for decisions at the level of the individual. Therefore, we present the covariance matrix implied by most factor score estimates as well as the covariance matrix implied by the unit-weighted scale model. Comparing these model-implied covariance matrices with the empirical covariance matrix of the causal-formative indicators allows for an evaluation of model fit.

As Bollen and Diamantopoulos (2015) have shown, some criticism is related to the fact that causal-formative indicators are confounded with other formative indicators, as for example composite-formative indicators. Composite-formative indicators are not based on conceptual unity. Accordingly, the relationship between the conceptual unity of the formative indicators, unit-weighted scale length, and model fit was investigated. The inter-correlation of the formative indicators was regarded as a measure of their conceptual unity. It was found for the model of parallel measurements that the unit-weighted scale model can –in theory– reach conventional criteria of model fit for the SRMR without substantial conceptual unity of the indicators when the number of indicators is greater than 150. The overall results for the population could be generalized to the sample and to the congeneric model by means of a simulation study. Since unit-weighted scales typically comprise much less than 150 indicators, the results indicate that for typical applications of the unit-weighted scales model, conceptual unity of the formative indicators is a necessary requirement.

It was, moreover, shown by means of an empirical example based on STAI data, that when a factor model (based on reflective indicators) has a rather moderate fit, the corresponding fit of the unit-weighted scale model and the fit of the model implied by the regression factor score estimate was worse. Thus, treating the reflective indicators of a factor model as causal-formative indicators in computing the unit-weighted scale and the factor score estimator resulted in an inacceptable model fit. The missing superiority of the factor score estimator over the unit-weighted scale could be due to the rather moderate fit of the corresponding CFA model for the STAI.

**Conclusion**

It was shown that factor score estimates as well as unit-weighted scales imply models that are based on causal-formative indicators. Factor models that are based on reflective indicators may be investigated to establish conceptual unity of indicators. However, factor models are not a sufficient basis for the validation of factor score estimates or unit-weighted scales, which are based on causal-formative indicators. It was therefore proposed to investigate the fit of the models implied by factor score estimates and unit-weighted scales. The model implied by unit-weighted scales fits to the data only when there is considerable conceptual unity of the indicators or when the number of indicators is extremely large. Therefore, conceptual unity of the causal-formative indicators is a prerequisite for unit-weighted scales, even when they are not based on factor analysis.

## Appendix

Syntax for SPSS Version 23.

```
MATRIX.

* Completely standardized factor loadings.
compute L={
  0.553  ;
  0.400  ;
  0.602  ;
  0.691  ;
  0.292  ;
  0.415  ;
  0.555  ;
  0.701  ;
  0.641  ;
  0.723  ;
  0.757  ;
  0.632  ;
  0.632  ;
  0.653  ;
  0.634  ;
  0.771  ;
  0.742  ;
  0.658  ;
  0.720  ;
  0.548   }.

* Inter-correlations between indicators (lower triangle).
* Indicators 1, 6, 7, 10, 13, 16, and 19 were reversed.
compute Sig={
1.000, 0.000, 0.000, 0.000, 0.000, 0.000, 0.000, 0.000, 0.000, 0.000, 0.000, 0.000, 0.000, 0.000, 0.000, 0.000, 0.000, 0.000, 0.000, 0.000;
0.159, 1.000, 0.000, 0.000, 0.000, 0.000, 0.000, 0.000, 0.000, 0.000, 0.000, 0.000, 0.000, 0.000, 0.000, 0.000, 0.000, 0.000, 0.000, 0.000;
0.295, 0.245, 1.000, 0.000, 0.000, 0.000, 0.000, 0.000, 0.000, 0.000, 0.000, 0.000, 0.000, 0.000, 0.000, 0.000, 0.000, 0.000, 0.000, 0.000;
0.457, 0.317, 0.447, 1.000, 0.000, 0.000, 0.000, 0.000, 0.000, 0.000, 0.000, 0.000, 0.000, 0.000, 0.000, 0.000, 0.000, 0.000, 0.000, 0.000;
0.130, 0.176, 0.072, 0.199, 1.000, 0.000, 0.000, 0.000, 0.000, 0.000, 0.000, 0.000, 0.000, 0.000, 0.000, 0.000, 0.000, 0.000, 0.000, 0.000;
0.279, 0.410, 0.194, 0.304, 0.147, 1.000, 0.000, 0.000, 0.000, 0.000, 0.000, 0.000, 0.000, 0.000, 0.000, 0.000, 0.000, 0.000, 0.000, 0.000;
0.278, 0.227, 0.426, 0.283, 0.017, 0.429, 1.000, 0.000, 0.000, 0.000, 0.000, 0.000, 0.000, 0.000, 0.000, 0.000, 0.000, 0.000, 0.000, 0.000;
0.323, 0.314, 0.421, 0.505, 0.192, 0.320, 0.418, 1.000, 0.000, 0.000, 0.000, 0.000, 0.000, 0.000, 0.000, 0.000, 0.000, 0.000, 0.000, 0.000;
0.261, 0.286, 0.410, 0.369, 0.325, 0.189, 0.381, 0.430, 1.000, 0.000, 0.000, 0.000, 0.000, 0.000, 0.000, 0.000, 0.000, 0.000, 0.000, 0.000;
0.575, 0.244, 0.343, 0.550, 0.180, 0.345, 0.326, 0.478, 0.350, 1.000, 0.000, 0.000, 0.000, 0.000, 0.000, 0.000, 0.000, 0.000, 0.000, 0.000;
0.379, 0.260, 0.423, 0.500, 0.193, 0.216, 0.415, 0.509, 0.551, 0.509, 1.000, 0.000, 0.000, 0.000, 0.000, 0.000, 0.000, 0.000, 0.000, 0.000;
```

```
0.356, 0.257, 0.392, 0.454, 0.306, 0.157, 0.310, 0.403, 0.419, 0.454, 0.578, 1.000, 0.000, 0.000, 0.000, 0.000, 0.000, 0.000, 0.000, 0.000;
0.429, 0.198, 0.236, 0.491, 0.141, 0.285, 0.315, 0.501, 0.290, 0.617, 0.426, 0.390, 1.000, 0.000, 0.000, 0.000, 0.000, 0.000, 0.000, 0.000;
0.281, 0.308, 0.340, 0.409, 0.239, 0.204, 0.328, 0.406, 0.539, 0.404, 0.552, 0.481, 0.384, 1.000, 0.000, 0.000, 0.000, 0.000, 0.000, 0.000;
0.465, 0.313, 0.507, 0.464, 0.152, 0.224, 0.283, 0.540, 0.370, 0.480, 0.474, 0.346, 0.301, 0.392, 1.000, 0.000, 0.000, 0.000, 0.000, 0.000;
0.486, 0.209, 0.443, 0.559, 0.239, 0.338, 0.368, 0.546, 0.413, 0.704, 0.555, 0.502, 0.599, 0.441, 0.463, 1.000, 0.000, 0.000, 0.000, 0.000;
0.329, 0.330, 0.473, 0.459, 0.195, 0.248, 0.432, 0.505, 0.701, 0.500, 0.650, 0.414, 0.398, 0.569, 0.462, 0.496, 1.000, 0.000, 0.000, 0.000;
0.365, 0.138, 0.525, 0.450, 0.266, 0.253, 0.349, 0.378, 0.403, 0.459, 0.507, 0.420, 0.360, 0.522, 0.385, 0.556, 0.540, 1.000, 0.000, 0.000;
0.393, 0.371, 0.432, 0.463, 0.192, 0.414, 0.600, 0.518, 0.452, 0.487, 0.537, 0.428, 0.536, 0.447, 0.419, 0.541, 0.495, 0.405, 1.000, 0.000;
0.141, 0.219, 0.418, 0.398, 0.182, 0.202, 0.380, 0.515, 0.367, 0.291, 0.409, 0.294, 0.276, 0.377, 0.396, 0.404, 0.383, 0.355, 0.443, 1.000
}.

* Compute a full corelation matrix.
compute Sig=Sig+T(Sig)-ident(nrow(sig),ncol(Sig)).

* Equation 7.
compute Sig_fs=L*INV(T(L)*INV(Sig)*L)*T(L).

* Equation 8.
compute SRMR_fs=( ( MSSQ(Sig - Sig_fs)+MSSQ(Diag(Sig - Sig_fs)) ) /(nrow(Sig)*(nrow(Sig)+1)) )&**0.5.

* Compute a unit-vector.
compute one=make(nrow(Sig),1,1).

* Equation 10.
compute Sig_tr=Sig*One*INV(T(one)*Sig*one)*T(one)*Sig.
compute SRMR_tr=( ( MSSQ(Sig - Sig_tr)+MSSQ(Diag(Sig - Sig_tr)) ) /(nrow(Sig)*(nrow(Sig)+1)) )&**0.5.

print {SRMR_fs, SRMR_tr}/format=F7.4.

END MATRIX.
```